\documentclass{article}

     \usepackage[final]{neurips_2019_ml4ps}

\usepackage[utf8]{inputenc} 
\usepackage[T1]{fontenc}    
\usepackage{hyperref}       
\usepackage{url}            
\usepackage{booktabs}       
\usepackage{amsfonts}       
\usepackage{nicefrac}       
\usepackage{microtype}  
\usepackage{mathtools}
\usepackage[pdftex]{graphicx}
\usepackage[ruled,vlined,noend]{algorithm2e}
\usepackage{fancyhdr,graphicx,amsmath,amssymb,amsfonts}
\usepackage[noend]{algpseudocode}
\usepackage[compact]{titlesec}
\usepackage[font=small]{caption}
\usepackage{url}
\usepackage{xcolor}
\usepackage[numbers]{natbib}

\title{A Taylor Based Sampling Scheme for Machine Learning in Computational Physics}

\author{Paul Novello\thanks{INRIA Paris Saclay, \{paul.novello,pietro.congedo\}@inria.fr}       \thanks{CEA CESTA, \{paul.novello, gael.poette, david.lugato\}@cea.fr}   \and \textbf{Gaël Poëtte}\footnotemark[2] \and \textbf{David Lugato}\footnotemark[2]  \and \textbf{Pietro Marco Congedo}\footnotemark[1]}

\begin{document}

\maketitle
\begin{abstract}
 Machine Learning (ML) is increasingly used to construct surrogate models for physical simulations. We take advantage of the ability to generate data using numerical simulations programs to train ML models better and achieve accuracy gain with no performance cost. We elaborate a new data sampling scheme based on Taylor approximation to reduce the error of a Deep Neural Network (DNN) when learning the solution of an ordinary differential equations (ODE) system.
\end{abstract}
\section{Introduction }

Computational physics allow simulating various physical phenomena when real measures and experiments are very difficult to perform and even sometimes not affordable. These simulations can themselves come with a prohibitive computational cost so that very often they are replaced by surrogate models \cite{Sudret,Karniadakis}. Recently, Machine Learning (ML) has proven its efficiency in several domains, and share similarities with classical surrogate models. The question of replacing some costly parts of simulations code by ML models thus becomes relevant. To reach this long term objective, this paper tackles a first step by investigating new ways to increase ML models accuracy for a fixed performance cost.

To this end we leverage the ability to control the sampling of the training data. Several previous works have hinted towards the importance of the training set, in the context of Deep Learning \cite{Bengio:2009:CL:1553374.1553380}, Active Learning \cite{series/synthesis/2012Settles} and Reinforcement Learning \cite{NIPS2010_3922}.
Our methodology relies on adapting the data generation procedure to gain accuracy with a same ML model exploiting the information coming from the derivatives of the quantity of interest w.r.t the inputs of the model. This approach is similar to Active learning, in the sense that we adapt our training strategy to the problem, but still differs because it is prior to the training of the model. Therefore, it is complementary with active learning methods (see \cite{series/synthesis/2012Settles} for a review and \cite{Gal,Zimmer} for more recent applications). The efficiency of this original methodology is tested on the approximation by a Deep Neural Network (DNN) of a stiff Bateman Ordinary Differential Equation (ODE) system. Solving this system at a moderate cost is essential in many physical simulations (neutronic \cite{gaek,Dufek}, combustion \cite{Bisi}, detonic \cite{LUCOR}, etc.). 

\section{Methodology}

To introduce the general methodology, we consider the problem of approximating a function $f:\mathbf{S} \subset \mathbb{R}^{n_i} \rightarrow \mathbb{R}^{n_o}$ where $\mathbf{S}$ is a subspace defined by the physics of interest. 
Let a model $f_{\theta}$, whose parameters $\theta$ have to be found in order to minimize a loss function $L(\theta) = \|f - f_{\theta}\|$. In supervised learning, we are usually given a training data set of $N$ points, $\{X_1, ... , X_N\}$, and their pointwise values $\{f(X_1),...,f(X_N)\}$. These points allow to statistically estimate $L(\theta)$ and then to use optimization algorithms to find a minimum of $L(\theta)$ w.r.t. $\theta$. \\
Amongst ML techniques, we chose $f_{\theta}$ to be a DNN for two reasons. First, DNNs outperform most of other ML models in high dimensions i.e. $n_i , n_o \gg 1$ which is often true in computational physics. Second, recent advances in Deep Learning frameworks have made DNNs much more efficient. Better optimization algorithms as well as the compatibility with GPUs and TPUs have greatly increased its performances.

\textbf{Sampling hypothesis} - The methodology is based on the assumption that a given DNN yields better accuracy when the dataset focuses on regions where $f$ is locally steeper. To identify these regions, we make use of the Taylor approximation (multi-index notation) for order $n$ on $f$:\\
\begin{minipage}{0.55\textwidth}
    \begin{equation}
        f(X+\boldsymbol{\epsilon}) \underset{\mathrm{\|\boldsymbol{\epsilon} \| \rightarrow 0}}{=} \sum_{0 \leq |\boldsymbol{k}| \leq n}  \boldsymbol{{\epsilon}^k} \frac{\partial^{\boldsymbol{k}}f(X)}{\boldsymbol{k}!} + O(\boldsymbol{\epsilon^{n}}). 
        \label{taylor}
    \end{equation}
\end{minipage}
\begin{minipage}{0.45\textwidth}
    \begin{equation}
        D^{n}_{\boldsymbol{\epsilon}}(X) =  \sum_{1 \leq |\boldsymbol{k}| \leq n}  \boldsymbol{{\epsilon}^k} \frac{\|\partial^{\boldsymbol{k}}f(X)\|}{\boldsymbol{k}!}.
        \label{dn}
    \end{equation}
\end{minipage}


Quantity $f(X+ \boldsymbol{\epsilon}) - f(X)$ derived using \eqref{taylor} gives an indication of how much $f$ changes around $X$. By neglecting the orders above $\boldsymbol{\epsilon^{n}}$, it is then possible to find the regions of interest by focusing on $D^{n}_{\boldsymbol{\epsilon}}$, given by \eqref{dn}. Notice that $D^{n}_{\boldsymbol{\epsilon}}$ is evaluated using $\|\partial^{\boldsymbol{k}}f(X)\|$ instead of $\partial^{\boldsymbol{k}}f(X)$ for derivatives not to cancel each other. The next steps are to evaluate and sample from $D^{n}_{\boldsymbol{\epsilon}}$.  

\textbf{Evaluating $ \boldsymbol{D^{n}_{\boldsymbol{\epsilon}}(x)}$} - \eqref{dn} involves the computation of derivatives of $f$. Usually in supervised learning, only $\{f(X_1),...,f(X_N)\}$ are provided and the derivatives of $f$ are unknown. However, here the dataset is drawn from a numerical simulation software. It is therefore possible either to use finite difference to approximate the derivatives, or to compute them exactly using automatic differentiation if we have access to the implementation. In any case, $\{\partial^{\boldsymbol{k}}f(X_1),..., \partial^{\boldsymbol{k}}f(X_N)\}$, and then $\{D^{n}_{\boldsymbol{\epsilon}}(X_1),...,D^{n}_{\boldsymbol{\epsilon}}(X_N)\}$ can be computed along with $\{f(X_1),...,f(X_N)\}$.

\textbf{Sampling procedure} - According to the previous assumption, we want to sample more where $D^{n}_{\boldsymbol{\epsilon}}$ is higher. To this end, we can build a probability density function (pdf) from $D^{n}_{\boldsymbol{\epsilon}}$, which is possible since $D^{n}_{\boldsymbol{\epsilon}} \geq 0$. It remains to normalize it but in practice it is enough considering a distribution $d \propto D^{n}_{\boldsymbol{\epsilon}}$. Here, to approximate $d$ we use a Gaussian Mixture Model (GMM) with pdf $d_{\text{GMM}}$ that we fit to $\{D^{n}_{\boldsymbol{\epsilon}}(X_1),...,D^{n}_{\boldsymbol{\epsilon}}(X_N)\}$ using the Expectation-Maximization (EM) algorithm.  $N'$ new data points $\{\bar{X}_1,...,\bar{X}_{N'}\}$, can be sampled, with $\bar{X} \sim d_{\text{GMM}}$. Finally, using the simulation software, we obtain $\{f(\bar{X}_1),...,f(\bar{X}_{N'})\}$, add it to $\{f({X}_1),...,f({X}_{N})\}$ and train our DNN on the whole dataset.  

\textbf{Methodology recapitulation} - Our methodology, which we call Taylor Based Sampling (TBS) is recapitulated in Algorithm \ref{construct}. \textbf{Line 1:} The choices of $\boldsymbol{\epsilon}$, the number of Gaussian distribution $n_{\text{GMM}}$ and $N'$ are not mandatory at this stage. Indeed, they are not a prerequisite to the computation of $\partial^{\boldsymbol{k}}f(x)$, which is the computationally costly part of evaluating $D^{n}_{\boldsymbol{\epsilon}}$. It allows to choose parameters $\boldsymbol{\epsilon}$ and $n_{\text{GMM}}$ \textit{a posteriori}. In this work, our choice criterion is to avoid sparsity of $\{\bar{X}_{1},...,\bar{X}_{N'}\}$ over $\mathbf{S}$. We use the Python package \texttt{scikit-learn} \cite{scikit-learn}, and more specifically the \texttt{GaussianMixture} class. \textbf{Line 2:} Usually in physical simulations, the input subspace $\mathbf{S}$ is bounded. Without \textit{a priori} informations on $f$, we sample the first points uniformly in a subspace $\mathbf{S}$. \textbf{Line 3-4:} To compute the derivatives of $f$ and because we have access to its implementation, we use the python package \texttt{jax} \cite{jax}, which allows automatic differentiation of \texttt{numpy} code. \textbf{Line 7-13:} Because the support of a GMM is not bounded, some points can be sampled outside $\mathbf{S}$. We recommend to discard these points and sample until all points are inside $\mathbf{S}$. This rejection method is equivalent to sampling points from a truncated GMM. 

\LinesNumbered
\setlength{\intextsep}{10pt}
\renewcommand{\thealgocf}{}
\begin{algorithm}
\SetAlgoLined
\textbf{Require: }$\boldsymbol{\epsilon}$, $N$, $N'$, $n_{\text{GMM}}$, $n$\\
Sample $\{X_1, ... , X_N\}$, with $X \sim \mathcal{U}(\mathbf{S})$ \\
\For(\Comment{\CommentSty{Note that order $k=0$ builds $\{f({X}_{1}),...,f({X}_{N})\}$}}){$0 \leq k \leq n$}{
Compute $\{\partial^{\boldsymbol{k}}f(X_1),...,\partial^{\boldsymbol{k}}f(X_N)\}$ using the simulation software.
}
Compute $\{D^{n}_{\boldsymbol{\epsilon}}(X_1),...,D^{n}_{\boldsymbol{\epsilon}}(X_N)\}$ using \eqref{dn}\\
Approximate $d  \sim D_{\boldsymbol{\epsilon}}$ with a GMM using EM algorithm to obtain a density $d_{\text{GMM}}$\\
$i \leftarrow 1$ \\
\While(\Comment{\CommentSty{Rejection method to prevent EM to sample outside $\mathbf{S}$}}){$i \leq N'$}{
    Sample  $\bar{X}_{i} \sim d_{\text{GMM}}$\\
    \uIf{$\bar{X}_{i} \notin \mathbf{S}$}
    {discard $\bar{X}_{i}$}
    \Else{
    $i \leftarrow i+1$}
}
Compute $\{f(\bar{X}_{1}),...,f(\bar{X}_{N'})\}$ \\
Add $\{f(\bar{X}_{1}),...,f(\bar{X}_{N'})\}$ to $\{f({X}_{1}),...,f({X}_{N})\}$
\caption{Taylor Based Sampling (TBS)}
\label{construct}
\end{algorithm}

\section{Application to an ODE system}
We apply our method to the resolution of the Bateman equations, which is an ODE system : 

\begin{equation*}
\begin{dcases}
    \partial_t u(t)   &= v\boldsymbol{\sigma_a} \cdot \boldsymbol{\eta}(t)u(t), \\
    \partial_t\boldsymbol{\eta}(t) &= v\boldsymbol{\Sigma_r} \cdot  \boldsymbol{\eta}(t)u(t),  \\
\end{dcases}
\text{,    with initial conditions    }
\begin{dcases}
    u(0) = u_0, \\
    \boldsymbol{\eta}(0) = \boldsymbol{\eta_0}. \\ 
\end{dcases},
\end{equation*}

with $u \in \mathbb{R}^+, \boldsymbol{\eta}\in(\mathbb{R}^{+})^M,\boldsymbol{\sigma}_a^T\in\mathbb{R}^M,\boldsymbol{\Sigma}_r \in\mathbb{R}^{M\times M}$. Here, $f: (u_0, \boldsymbol{\eta_0}, t) \rightarrow (u(t), \boldsymbol{\eta}(t))$.\\
For physical applications, $M$ ranges from tens to thousands. 
We consider the particular case $M=1$ so that $f:\mathbb{R}^3 \rightarrow \mathbb{R}^2$, with $f(u_0, \eta_0, t) = (u(t), \eta(t))$. The advantage of $M=1$ is that we have access to an analytic, cheap to compute solution for $f$. It allows to conduct extensive analyses for the design of our methodology. Of course, this particular case can also be solved using a classical ODE solver, which allows us to test it end to end. It can thus be generalized to higher dimensions ($M > 1$).\\
All DNN trainings have been performed in \texttt{Python}, with \texttt{Tensorflow} \cite{Tensorflow}. We used a fully connected DNN with hyperparameters chosen using a simple grid search. The final values are: 2 hidden layers, "ReLU" activation function, and 32 units for each layer, trained with the Mean Squared Error (MSE) loss function using Adam optimization algorithm with a batch size of 50000, for 40000 epochs and on $N+N' = 50000$ points, with $N=N'$. We first trained the model for $(u(t), \eta(t)) \in \mathbb{R}$, with an uniform sampling, that we call basic sampling (BS) ($N'=0$), and then with TBS for several values of $n$, $n_{\text{GMM}}$ and $\boldsymbol{\epsilon}=\epsilon(1,1,1)$, to be able to find good values. We finally select  $\epsilon = 5\times 10^{-4}$, $n=2$ and $n_{\text{GMM}} = 10$. The data points used in this case have been sampled with an explicit Euler scheme. This experiment has been repeated 50 times to ensure statistical significance of the results. 

\section{Results and discussion} 

\textbf{Table 1} summarizes the MSE, i.e. the $L_2$ norm of the error of $f_{\theta}$ and $L_{\infty}$ norm, with $L_{\infty}(\theta) = \underset{X\in \mathbf{S}}{\max}(|f(X) - f_{\theta}(X)|)$ obtained at the end of the training phase. This last metric is important because the goal in computational physics is not only to be averagely accurate, which is measured with MSE, but to be accurate over the whole input space $\mathbf{S}$. Those norms are estimated using a same test data set of $N_{test} = 50000$ points. The values are the means of the $50$ independent experiments displayed with a $95\%$ confidence interval. These results reflect an error reduction of 6.6\% for $L_2$ and of 45.3\% for $L_{\infty}$, which means that TBS mostly improves the $L_{\infty}$ error of $f_{\theta}$. Moreover, the $L_{\infty}$ error confidence intervals do not intersect so the gain is statistically significant for this norm.

\begin{table}[h]
  \caption{Comparison between \textcolor[rgb]{0.137,0.341,0.537}{BS} and \textcolor[rgb]{0.85,0.282,0}{TBS}}
  \label{sample-table}
  \centering
  \begin{tabular}{lllll}
    \toprule
    \cmidrule(r){1-5}
     Sampling & $L_2$ error $ (\times 10^{-4})$ & $L_{\infty}$ $ (\times 10^{-1})$ & AEG$ (\times 10^{-2})$ & AEL$ (\times 10^{-2})$\\
    \midrule
    BS & $1.22 \pm 0.13 $ & $5.28 \pm 0.47$ & - & -    \\
    \textbf{TBS}     & $\boldsymbol{1.14} \pm 0.15 $ & $\boldsymbol{2.96} \pm 0.37 $  &  $2.51 \pm 0.07 $ & $0.42 \pm 0.008 $  \\
    \bottomrule
  \end{tabular}
\end{table}

\textbf{Figure 1a} shows how the DNN can perform for an average prediction. \textbf{Figure 1b} illustrates the benefits of \textcolor[rgb]{0.85,0.282,0}{TBS} relative to \textcolor[rgb]{0.137,0.341,0.537}{BS} on the $L_{\infty}$ error (Figure 2b). These 2 figures confirm the previous observation about the gain in $L_{\infty}$ error.   
Finally, \textbf{Figure 2a} displays $u_0,\eta_0 \rightarrow \underset{0 \leq t \leq 10} {\max} D^{n}_{\boldsymbol{\epsilon}}(u_0, \eta_0, t)$ w.r.t. $(u_0, \eta_0)$ and shows that $D^{n}_{\boldsymbol{\epsilon}}$ increases when $U_0 \rightarrow 0$. TBS hence focuses on this region. Note that for the readability of this plots, the values are capped to $0.10$. Otherwise only few points with high $D^{n}_{\boldsymbol{\epsilon}}$ are visible. \textbf{Figure 2b} displays $u_0,\eta_0 \rightarrow g_{\theta_{BS}}(u_0,\eta_0) -g_{\theta_{TBS}}(u_0,\eta_0)$, with  $g_{\theta}:u_0,\eta_0 \rightarrow \underset{0 \leq t \leq 10} {\max}\|f(u_0,\eta_0,t) - f_{\theta}(u_0,\eta_0,t)\|_2^2$ where $\theta_{BS}$ and $\theta_{TBS}$ denote the parameters obtained after a training with BS and TBS, respectively.  It can be interpreted as the error reduction achieved with TBS. The highest error reduction occurs in the expected region. Indeed more points are sampled where $D^{n}_{\boldsymbol{\epsilon}}$ is higher. The error is slightly increased in the rest of $\mathbf{S}$, which could be explained by a sparser sampling on this region. However, as summarized in \textbf{Table 1}, the average error loss (AEL) of TBS is around six times lower than the the average error gain (AEG), with $AEG = \mathbb{E}_{u_0,\eta_0}(Z\mathbb{1}_{Z>0})$ and $AEL = \mathbb{E}_{u_0,\eta_0}(Z\mathbb{1}_{Z<0})$ where $Z(u_0,\eta_0) =  g_{\theta_{BS}}(u_0,\eta_0) -g_{\theta_{TBS}}(u_0,\eta_0)$. In practice, AEG and AEL are estimated using uniform grid integration, and averaged on the $50$ experiments. 

\begin{figure}[!htb]
   \begin{minipage}{0.5\textwidth}
     \includegraphics[trim={0cm 0cm 0cm 0cm},clip,width=1.0\linewidth]{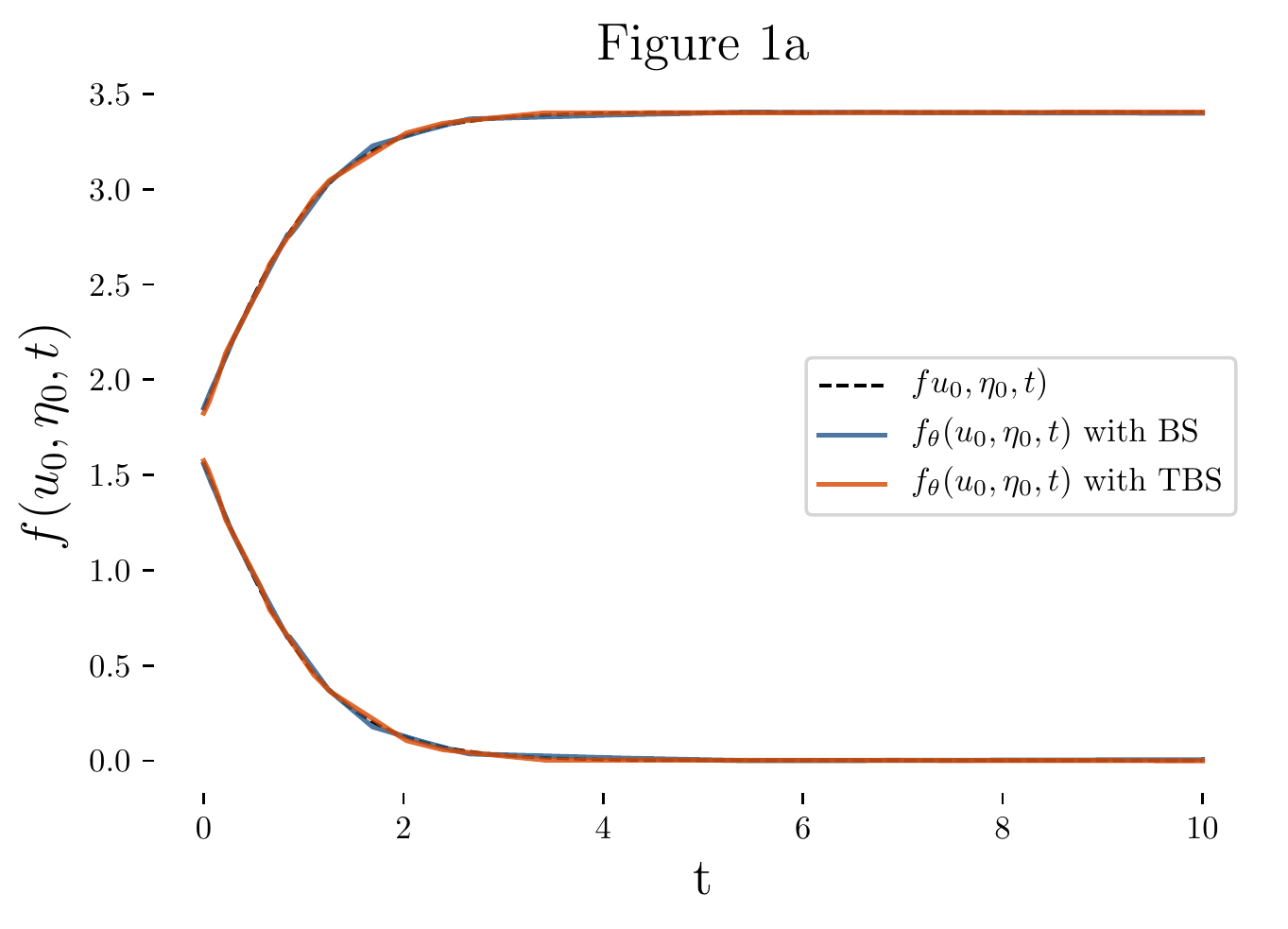}
     \vspace{-0.5cm}
   \end{minipage}\hfill
   \begin{minipage}{0.5\textwidth}
     \includegraphics[trim={0cm 0cm 0cm 0cm},clip,width=1.0\linewidth]{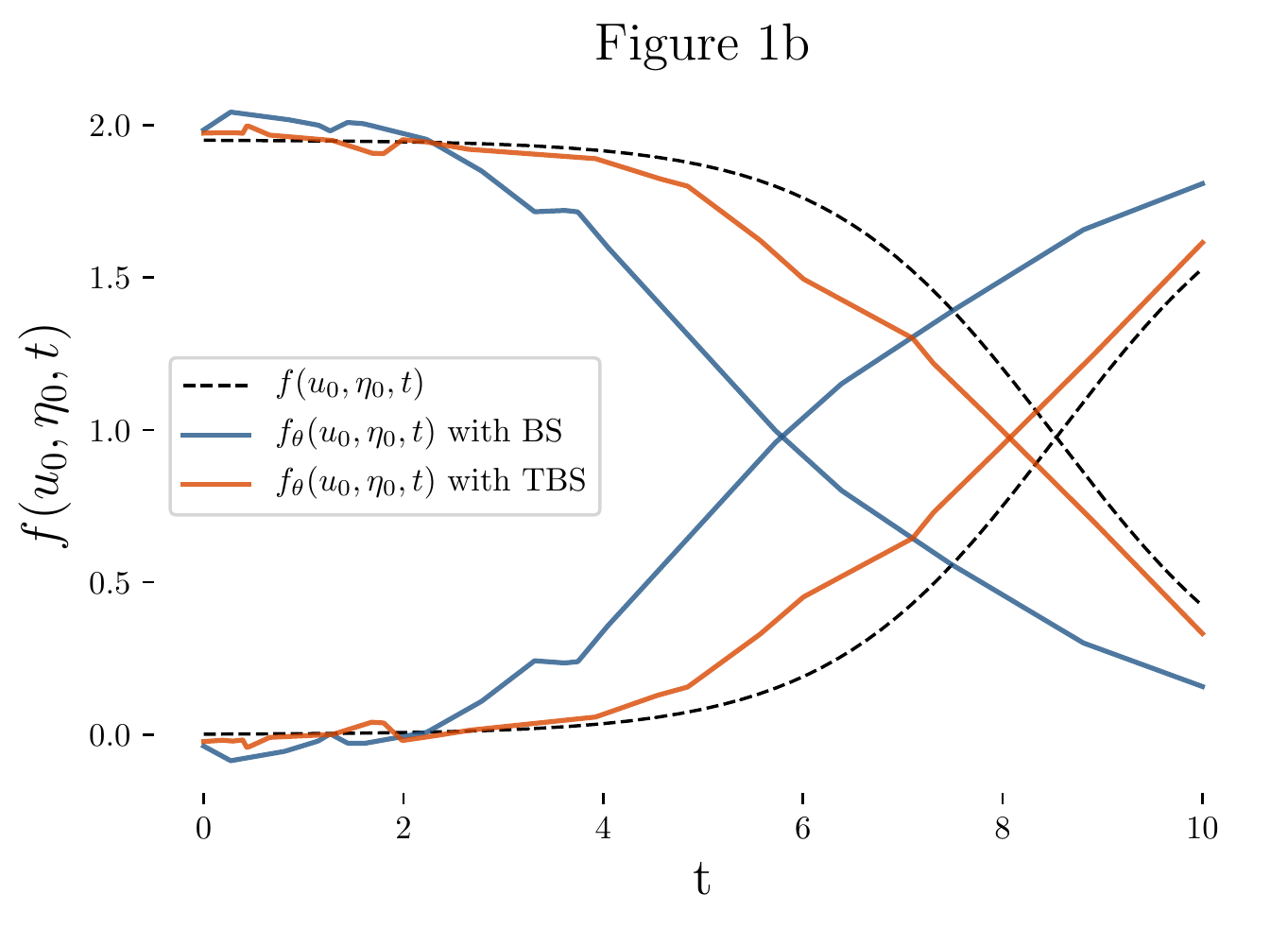}
     \vspace{-0.5cm}
   \end{minipage}
\end{figure}
\begin{figure}[!htb]
   \begin{minipage}{0.5\textwidth}
     \centering
     \includegraphics[trim={0cm 0cm 0cm 0cm},clip,width=1.01\linewidth]{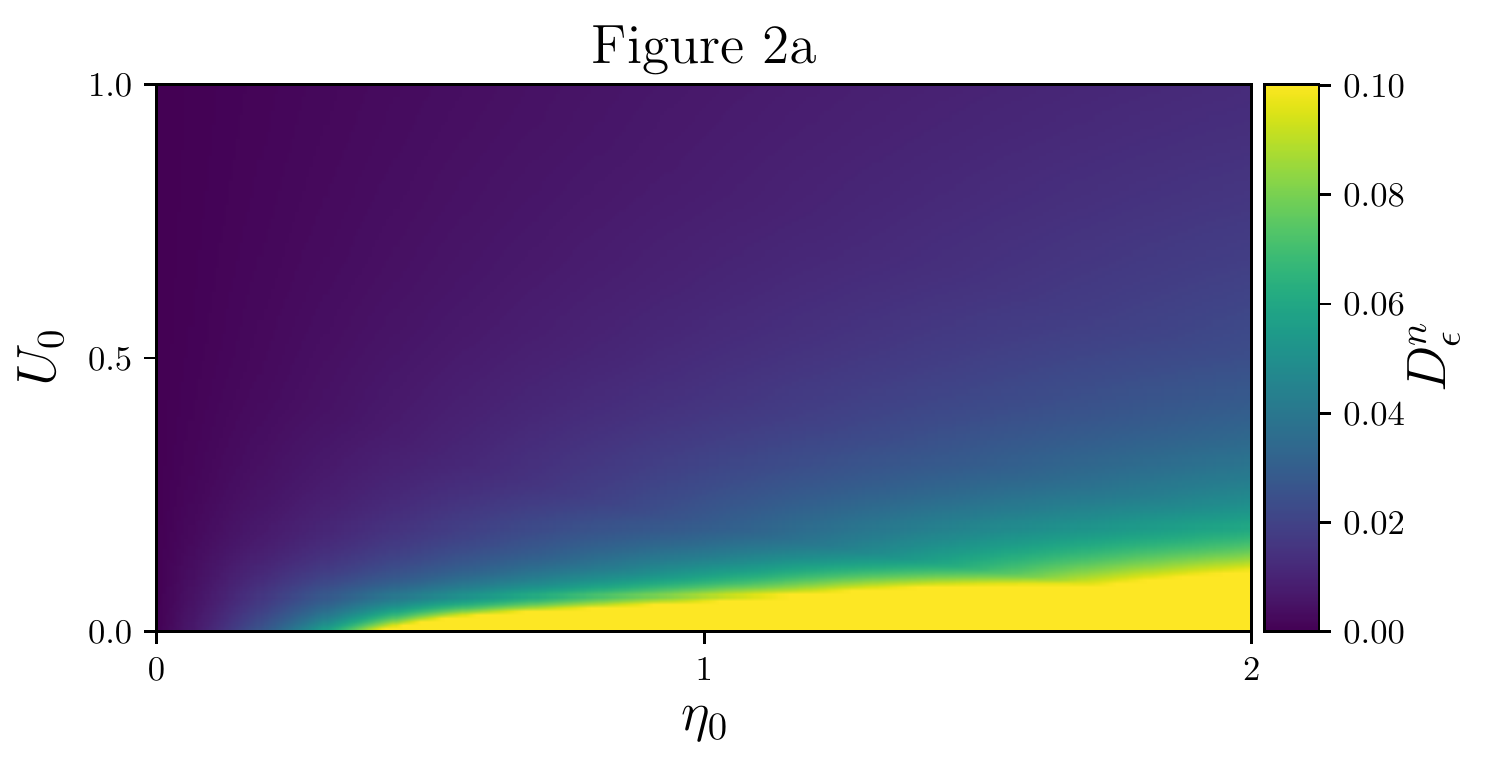}
   \end{minipage}\hfill
  \begin{minipage}{0.5\textwidth}
     \centering
     \includegraphics[trim={0cm 0cm 0cm 0cm},clip,width=1.01\linewidth]{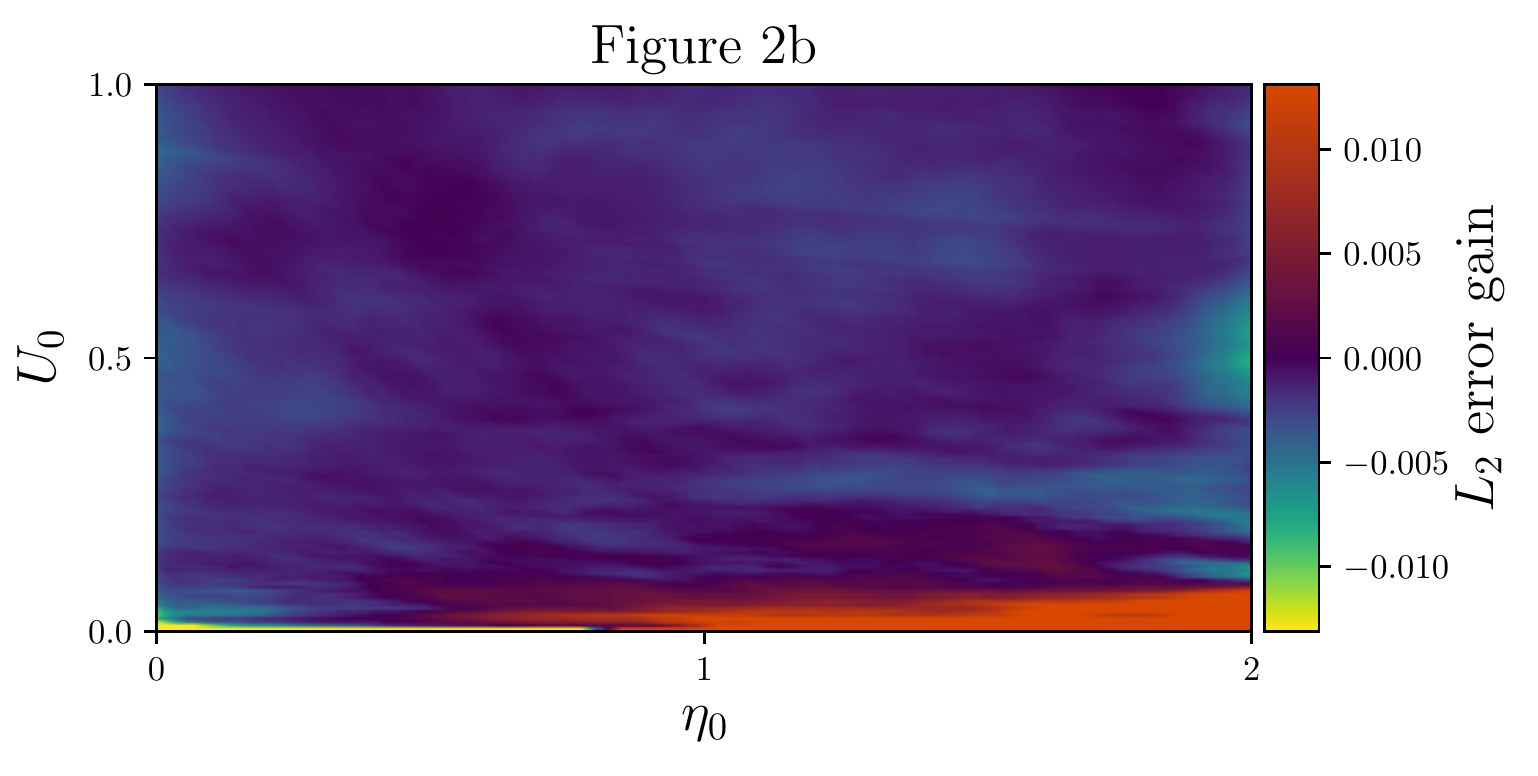}
   \end{minipage}
   \caption*{\textbf{1a: } $t \rightarrow f_{\theta}(u_0,\eta_0,t)$ for randomly chosen $(u_0,\eta_0)$, for $f_{\theta}$ obtained with the two samplings. \textbf{1b:} $t \rightarrow f_{\theta}(u_0,\eta_0,t)$ for $(u_0,\eta_0)$ resulting in the highest point-wise error with the two samplings. \textbf{2a:} $u_0,\eta_0 \rightarrow \underset{0 \leq t \leq 10} {\max} D^{n}_{\boldsymbol{\epsilon}}(u_0, \eta_0, t)$ w.r.t. $(u_0, \eta_0)$. \textbf{2b:} $u_0,\eta_0 \rightarrow g_{\theta_{BS}}(u_0,\eta_0) -g_{\theta_{TBS}}(u_0,\eta_0)$, }
\end{figure}

Results are promising for this case of the Bateman equations ($M = 1$). We selected a simple numerical simulation to efficiently test the initial assumption and elaborate our methodology. Its application to more expensive physical simulations is the next step of this work. By then, several problems have to be tackled. 
First, TBS does not scale well to higher dimensions, because it involves the computations of $|\partial^{\boldsymbol{k}}f(x)|$, i.e. $n_o \times n_i^{n+1}$ derivatives for each $D^{n}_{\boldsymbol{\epsilon}}$. This issue is less important when using automatic differentiation than finite difference, and we can compute the derivatives for only few points (i.e. chose $N < N'$) to ease it, but we will investigate on how to traduce the initial assumption for high dimensions. Moreover, the exploration of the input space is more difficult in this case, because the initial sampling is more sparse for a same $N$ when $n_i$ increases. 
In this work, $\boldsymbol{\epsilon}$, $N'$, $n$ have only been empirically chosen, and we arbitrarily selected a GMM to approximate $d$. These choices have to be questioned in order to improve the methodology.
Finally, this paper focused on accuracy rather than performance, whereas performance is the goal of using a surrogate in place of a simulation code. Extending TBS to higher dimensions will allow to investigate this aspect. Beside, DNN may not be the best ML model in every situation, in terms of performances, but the advantage of TBS is that it can be applied to any ML model.

\section{Conclusion}

We described a new approach to sample training data based on a Taylor approximation in the context of ML for approximation of physical simulation codes. Though non specific to Deep Learning, we applied this method to the approximation of the solution of a physical ODE system by a Deep Neural Network and increased its accuracy for a same model architecture. \\
In addition to the leads mentioned above, the idea to use the derivatives of numerical simulations to better train ML models should be explored. This idea has already been investigated in other fields such as gradient-enhanced kriging \cite{gradient-krigging} or Hermite interpolation \cite{hermite} and could lead to ML approaches based on higher orders. An example of application could be to include the derivatives as new training points and making a DNN learn these derivatives.

\small
\bibliographystyle{plain} 
\bibliography{refs}

\end{document}